\documentclass[aps,pra,preprint,groupedaddress]{revtex4}

\newcommand{\be}{\begin{equation}}
\newcommand{\ee}{\end{equation}}

\begin{document}

\title{
Anomalous low temperature specific heat of $^3$He inside nanotube bundles}

\author{ S.M. Gatica$^1$, F. Ancilotto$^2$,   and  M.W. Cole$^1$}

\affiliation{
$^1$  Department  of
Physics, Pennsylvania State
University, University Park, PA 16802\\
$^2$INFM (UdR Padova and DEMOCRITOS National Simulation Center, Trieste, Italy)  and Dipartimento di Fisica ``G. Galilei'',
Universit\`a di Padova,\\ via Marzolo 8, I-35131 Padova, Italy.}

\date{\today}

\begin{abstract}

Helium atoms and hydrogen molecules can be strongly bound inside interstitial channels within bundles of carbon nanotubes. An exploration of the low energy and low temperature properties of $^3$He atoms is presented here. Recent study of the analogous  $^4$He system has shown that the effect of heterogeneity is to yield a density of states $N(E)$ that is qualitatively different from the one-dimensional (1D) form of $N(E)$ that would occur for an ideal set of identical channels. In particular, the functional form of $N(E)$ is that of a 4D gas near the very lowest energies and a 2D gas at somewhat higher energies. Similar behavior is found here for $^3$He. The resulting thermodynamic behavior of this fermi system is computed, yielding an anomalous form of the heat capacity and its dependence on coverage.

\end{abstract}

\maketitle

\section{Introduction}

Extensive efforts have been made to study the essentially one-dimensional (1D) behavior of gases exposed to the very linear, confining geometries provided by ensembles of carbon nanotubes and other quasi-1D porous materials, such as zeolites and MCM-41 \cite{1,2}. Thermodynamic and structural measurements are traditional ways to explore such reduced dimensional behavior. One of the more direct signatures of the 1D character of the adsorbate should appear in the heat capacity $C(T)$ in the temperature $(T)$   regime where phonon excitations characterize the dynamics of the adsorbate at high density \cite{3,4}. In that case, the power d of $T$ entering the low $T$ relation $C(T) \sim T^d $ is a function of the effective dimensionality of the adsorbate: $d=1$ for linear confinement, $d= 2$ in a monolayer,  etc.\cite{5} Another manifestation of d involves the form of $C(T)$ in the classical Boltzmann regime of very low coverage and high $T$; there, the behavior for a system of N particles is

\be
\frac{C}{N k_B} = \frac{d}{2}		
\ee

In some cases, however, such seemingly straightforward descriptions of the effective dimensionality can be questioned, leading to more careful investigations, with surprises to be found. Such was the case in our recent study (henceforth called I) of particles moving in a system of interstitial channels (ICs) between three adjacent carbon nanotubes \cite{6,7,8}. This system had previously been studied extensively with idealized models of the adsorption geometry, which predict 1D properties of the system.\cite{5,6,7,8,9,10,11,12,13} In the new study, in contrast, it was found that the density of states for single particles ($^4$He or H$_2$) is drastically altered by heterogeneity; some heterogeneity is always present within  experimental samples of nanotube bundles \cite{14,15}. We found in I that the density of states $N(E)$ near the very lowest energy has the form:

\be
N(E)= a (E - E_{min}) \Theta(E - E_{min}) \hspace{2cm}	(E \sim E_{min})	
\ee
Here $E_{min}$ is the lowest energy level present in the set of ICs, $\Theta(x)$ is the Heaviside unit step function and a is a constant that depends on both the experimental sample and the details of the dispersion relation of the particles. 

This linear dependence of $N(E)$ on $(E - E_{min})$ is qualitatively distinct from the 
$(E - E_{min})^{-1/2}$ dependence found for a conventional 1D gas. Eq. (2) arises (as discussed below) when the distribution of transverse energies associated with the heterogeneity is convoluted with the 1D density of states describing motion parallel to the tubes' axes. The behavior of $N(E)$ at very high energy is given instead by the relation expected for a uniform, homogeneous ensemble of $N_{IC}$ identical ICs:

\be
N_{homo}(E)= N_{IC} \frac{L}{(\hbar\pi)} \left[\frac{m}{2(E - E_{min})}\right]^{1/2}		(E \gg E_{min} ) 
\ee
Here, $L$ is the length of each nanotube (assumed constant) and $m$ is the particle mass. Eq. 3 describes a strictly 1D spectrum  of particles. In paper I we found that between the two energy regimes described by Eqs. (2) and (3) there exists another regime in which the density of states is nearly constant: $N(E) \sim (E - E_{min})^0$. Thus, the power $\lambda$ characterizing the energy dependence $N(E)\sim (E-E_{min})^{\lambda}$ varies from $\lambda$=1 to $\lambda$=0 to $\lambda$=-1/2 as $E$ increases. Recall that the relationship upon which Eq. 1 is based presumes that the density of states is that of a d-dimensional ideal gas, for which $\lambda  =d/2 -1$. Hence, the behavior of $N(E)$ found in I for the nonuniform IC distribution implies that the {\em effective} dimensionality of the gas is d=4 at the lowest energy, d=2 at intermediate energy and d=1 at high energy. This variation leads to a corresponding T-dependent power law behavior of $C(T)$ as $T$ increases.

In I, we examined the consequences of this unusual spectrum for the very low $T$ behavior of an ideal bose gas. The calculated properties include a Bose-Einstein condensation (BEC), characterized by 4D thermal properties at both very low $T$ and close to the transition temperature, evolving into a 2D gas at somewhat higher $T$ and, eventually, to a 1D gas at very high $T$. The present paper is directed instead at calculating the properties of an ideal fermi gas in this environment, focusing on the low $T$ properties; for this system, of course, there is no transition analogous to BEC. However, the behavior of the fermi system is sensitive to the forms of $N(E)$ in the ultra-low $T$ and low $T$ regimes. These regimes exhibit 4D and 2D behavior, respectively, for the function $C(T)$.

The next section evaluates the properties of a fermi gas of $^3$He atoms exposed to this nanotube bundle. Qualitatively similar behavior is expected to occur for a gas of HD molecules, but we have not undertaken the corresponding calculations in that case. The mathematical description of $N(E)$ in both paper I and in Section 2 assumes that the ICs within the experimental sample include some having the lowest possible energies, near $E_{min}$. Section 3 considers an alternative situation, corresponding to the case when the particular sample of tubes and ICs does {\em not} include those with such low energies; this alternative case leads to quite different predictions for $C(T)$ from those discussed in Section 2. Section 4 summarizes our results and draws some conclusions about this interesting system.

\section{ Properties of the $^3$He gas}

The methodology employed in I is used here to compute $N(E)$ for individual $^3$He atoms (which we assume to interact only with the sorbent). The single particle states are calculated by assuming that the atoms move freely in the z direction (parallel to the tubes' axes); in this case, the energy is the sum of a transverse energy and a longitudinal energy. The transverse energy is derived by numerical solution of the Schrodinger equation for each specified combination of radii surrounding an IC. For a given IC, we define a vector ${\bf R} =( R_1, R_2  , R_3  )$ with components equal to the values of these neighboring radii. A particular experimental sample is represented as a cloud of points in $\bf R$ space. The function $f({\bf R})$ is defined as the density of points in this abstract space. Note that $f({\bf R})$ is an extensive variable, which integrates over {\bf R} to yield the number of ICs present in the experimental sample:

\be
N_{IC} = \int d{\bf R} f({\bf R})
\nonumber
\ee
The calculations in this paper employ the same model potential used in I for $^4$He, so the only difference in $N(E)$ arises from the smaller mass of the $^3$He. For single wall nanotubes of typical diameter 1.4 to 2 nm, the IC provides a strongly attractive potential that restricts the probability density to close proximity to the center of the IC. Because the atoms are so highly localized, the excited states of the transverse spectrum lie far above the ground state energy, which we denote $E_t({\bf R})$; hence, these excited states may be neglected in computing the properties of the system that are relevant to low temperature. The total energy $E({\bf R},p)$ of a molecule in a particular state is then a function of four "quantum numbers": the three radii and the momentum $(p)$ parallel to the IC direction.

\be
E({\bf R},p) = E_t({\bf R}) + \frac{p^2}{2m}		
\ee
The density of states at energy E is derived by adding up contributions from all of the ICs present in a given experimental sample:

\be
N(E) = \sum_p \int d{\bf R} f({\bf R}) \delta[E- E({\bf R},p)]		
\ee
Here, the sum involves the usual quasicontinuous distribution of momenta. By integrating in the usual way over this variable, one obtains an expression involving the transverse density of states, $g(E_t)$:

\be
N(E) = \frac{L}{\pi\hbar}(\frac{m}{2})^{1/2} \int_{E_{min}}^E dE_t\, g(E_t) \,[E- E_t]^{-1/2}	
\ee

\be
g(E) = \int d{\bf R } f({\bf R}) \delta[E - E_t({\bf R}_0)]
\ee
In a hypothetical uniform case, all of the tubes have a common radius (i.e. ${\bf R} = (R_0,R_0,R_0)={\bf R_0}$ ) 
so that $f({\bf R}) =N_{IC} \delta({\bf R}-{\bf R}_0)$ ; then $g(E)= N_{IC}\delta[E-E_t({\bf R}_0)]$ and the result of 
evaluating Eq. 6 coincides with the 1D solution, Eq. 3, with $E_t({\bf R}_0)$ replacing $E_{min}$. 
In practice, the singular threshold behavior of Eq. 3 is exquisitely sensitive to heterogeneity. 
As a consequence, the low energy behavior of $N(E)$ is dramatically altered from this 1D form.
 We omit the details of the calculation, which are discussed in paper I. Fig. 1 presents the density of states for $^3$He in a distribution of ICs identical to that used in I for $^4$He. There is no evidence of any singular behavior because the inverse square root singularity  in Eq. 3 is removed by the heterogeneity; in this case, $N(E)$ at very low E is seen in Fig. 1 to exhibit the same linear dependence  (Eq. 1) on energy above threshold as was found in I, although the numerical value of the coefficient $a$ differs, as do the values of $E_{min}$ (-446.79K for $^3$He)  and $R_{min}$ (8.71 \AA\ ). This similarity is expected because the only difference between this case and that explored in I is the mass. The 4D-like, 
linear dependence of $N(E)$ arises from convoluting the 1D longitudinal spectrum with a 
 function $g(E)$ that is 3D-like, derived from those ICs for which ${\bf R}$ lies within a 3D annular region in ${\bf R}$ space near near ${\bf R}_{min}$ (corresponding to the energy domain $[E, E+dE]$); see I for a detailed discussion of this point.

The thermal properties of the gas are computed (as in I) by ignoring interparticle interactions. This assumption lacks rigorous justification but it may be appropriate in the nanotube environment due to electrodynamic screening of the interatomic interaction \cite{16}. Also not considered is elastic screening, associated with the tubes' deformation, which has not been studied, to the best of our knowledge. In the analogous case of adsorption on a flat surface, these effects are usually small \cite{17,18,book}, but the present situation differs because the adsorbate wave function is localized closer to the nanotubes than it is to the planar surface. Indeed, we have found at T=0 that the interaction leads to a non-negligible dilation of the lattice of nanotubes at high particle density and a factor of 10 smaller dilation at low density \cite{19,20,21}.

With the neglect of all such interactions, the energy and specific heat are computed in the conventional way for an ideal fermi gas. For example, the energy of the system satisfies

\be
U= \int dE \,E  \, N(E)\,  \{\exp[\beta(E-\mu)] +1\}^{-1}	
\ee
Here $1/\beta=k_BT$ and $\mu$ is the chemical potential, which is determined from the equation for the total number of atoms:

\be
N= \int dE N(E) \{\exp[\beta(E-\mu)] +1\}^{-1}	
\ee
The classical limiting behavior corresponds to the case of low density, when 
$\exp[\beta (E-\mu)]\gg 1$. Then, the energy per particle is

\be
(\frac{U}{N})_{classical} = -\frac{d(ln Q)}{d\beta} 		
\ee
where the single particle partition function is

\be
Q= \int dE N(E) \exp[-\beta E].	
\ee
The Fermi energy corresponding to the low energy limiting form of the density of states, Eq. 2, is given by the relation

\be
E_f = \frac{2N}{a}^{1/2}	
\ee
Fig. 2 shows the behavior of $E_f$  over the energy regime extending to higher energy, the so-called 2D regime where $N(E)$ is constant.

The effective dimensionality of the system varies with both $N$ and $T$ since the low $T$ behavior probes the $E_f$ dependence of $N(E_f)$, which varies with N.  Figure 3 presents 
results obtained for $C(T) = (dU/dT)_N$ at very low $T$. The behavior at the lowest density 
corresponds closely to the classical limit ($N$ approaching 0), except for $T<$ 1mK, 
when degeneracy effects appear. That limit has the behavior described qualitatively 
in the introduction. The $T=0$ classical limit is $C/(N k_B) =2$, since the lowest energy 
density of states has the form ($N$ proportional to $E$) of a 4D gas. Between 1 mK and 15 mK the behavior is that of a classical gas with the density of states shown in Fig. 1. 
At higher $T$ (above 0.02 K), 
the result is $C/(N k_B) \simeq 1$, that of a 2D gas. At much higher $T$ (not shown), the 
classical specific 
heat becomes that of a 1D gas, $C/(N k_B)=1/2$,  since then the effects of heterogeneity are minimal.

This behavior evolves as the number of atoms, $N$, increases. At intermediate values of $N$,
 the degenerate regime of $C\sim T$ extends to higher $T$ before converging to the classical
 limit. At the highest density appearing in Fig. 3a, there is little evidence of the 4D 
behavior and the result is similar to that of a strictly 2D gas. This is demonstrated
 in Fig. 2 by an explicit comparison with the prediction for the 2D gas. Note that
 the low $T$ limiting behavior is $C/k_B= \alpha T$, where 
$\alpha = (\pi^2/3) k_B^2 N(E_f)$ , i.e., proportional to the density of states at
 the fermi level. When the density of states is linear in $E$, $\alpha$ is proportional 
to $N^{1/2}$. 
The general dependence can be undestood from Eqs. 1 and 2, which determine how $\alpha$ depends on $N$. 

\section{Alternative (gaussian) model calculation}

In the preceding discussion, the sample's distribution of tubes is  assumed to include ICs in the 
vicinity of the global energy minimum, $E_{min}$. In this section, we consider the extreme
 opposite case, in which the distribution function $f({\bf R})$ is centered about a 
region of ${\bf R}$ space near a point ${\bf R}_0$ that is far from ${\bf R}_{min}$. 
We assume a gaussian form for this distribution:

\be
f({\bf R}) = F \; exp[-({\bf R}- {\bf R}_0)^2/(2 \sigma^2)]	
\ee
Here $\sigma$ is the distribution's width parameter and $F= N_{IC} (2\pi)^{-3/2}\sigma^{-3}$ is a 
normalization constant, yielding the integral N$_{IC}$ for $f({\bf R})$. The transverse density of
 states is obtained from Eq. 8, which yields the relation \cite{footnote}

\be
g(E)= \int d^2 {\bf S} f({\bf R})/\mid\nabla E_t({\bf R})\mid	
\ee
Here, the integral is over the equal energy surface ${\bf S}$ in ${\bf R}$ 
space that satisfies $E= E_t({\bf R})$. In the following, we assume that $\sigma$ is "small",
 so that a linear expansion of the energy in the vicinity of $E_0 = E_t({\bf R}_0)$ suffices
 to characterize
 the transverse energy within the distribution:

\be
E_t({\bf R}) =  E_{t0} +  \nabla E_t\cdot ({\bf R}- {\bf R}_0)	
\ee
Here $E_{t0}= E_t({\bf R}_0)$ is the most probable transverse energy within the distribution of 
ICs and  $\nabla E_t$ is the gradient evaluated at $R_0$.With this expansion and the gaussian
 distribution 
function, Eq. 14 can be evaluated analytically, yielding

\be
g(E_t) = \frac{2 F \pi \sigma^2 }{\mid\nabla E_t \mid} 
\exp\{\frac{-[E-E_{t0}]^2 } {2 (\sigma \mid\nabla E_t\mid)^2} \}
\ee
As might have been expected, this distribution assumes a gaussian form. The standard deviation of the energy distribution 
$\sigma \mid\nabla E_t\mid$ is large if the cloud of IC points representing the sample is spread broadly 
in {\bf R} space and/or the energy gradient is large.

Our analytical result [17] for this gaussian model represents a stark contrast to the
 result $g(E_t) \sim (E_t- E_{min})^{3/2}$ found near the energy threshold in I, a consequence of 
the distribution of ICs being localized around ${\bf R}_{min}$ ; it is the 3/2 power law dependence that
 yields the d=4 low energy spectrum discussed in the previous section of this paper.
 In the present case, Eq. 17 may be integrated, using Eq. 7, to obtain the total $N(E)$. The result 
appears for two sets of parameters in Fig.4

One observes in these curves that the 1D character of $N(E)$ is preserved at
 high  $E$, but an exponential tail of low energy states appears, which will 
determine the low energy/$T$ behavior of the adsorbed gas. Fig 5 exhibits the 
heat capacity obtained with varying coverages with this model.

The most interesting feature of these results is the extreme sensitivity of 
the low energy and low $T$ behaviors to the distribution of ICs. At the very 
lowest energy, $N(E)$ is very flat, which might be interpreted as a very 
high effective dimensionality, meaning that C assumes a high value in the
 classical regime; see Eq. 1. The coefficient of the specific heat in the linear, 
degenerate regime is very sensitive to both the number of atoms present and 
the form of $N(E)$, as usual.

\section{Summary and conclusions}

In this paper we have explored the spectrum and thermal properties of 
a gas of $^3$He atoms confined within a set of ICs. Two distinct situations 
were considered. The first study assumed that the distribution of ICs in ${\bf R}$
 space is localized near the global energy minimum, as was assumed in I. 
In this case, the low $T$ (4D) behavior of $C(T)$ is very $N$ dependent, as seen in 
Fig. 3. There occurs a maximum in $C(T)$ during the crossover between 
the regimes of degenerate and classical statistics. One can, in principle,
 have 1D, 2D or 4D behavior in classical and quantum regimes. The description, 
in general, is complicated because the low $T$ behavior is determined by the number 
of particles, which determines the density of states at the fermi level. In the
 alternative gaussian model the low $T$ behavior is very different from that of
 the preceding model. The low $T$ behavior is determined by the tail of the 
exponential at low $N$, which does not correspond to any particular effective 
dimensionality. In contrast, at high $N$ (in both models) the properties are those
 of a 1D system.

This research has been supported by the National Science Foundation.
We are indebted to Mercedes Calbi, Carlo Carraro, Tony Clark, Susana Hern\'andez, Karl Johnson and  John Saunders 
  for discussions of this work.

\newpage

FIGURE CAPTIONS

1. The density of states of $^3$He atoms in the vicinity of the lowest energy $E_{min}$

2. Fermi energy as a function of $N$.

3. a)The specific heat of a gas of $^3$He atoms at densities $N=10^{-5}\AA^{-3}$, $N=10^{-6}\AA^{-3}$ and  $N=10^{-7}\AA^{-3}$.  
 b) Specific heat of an ideal 2D fermi gas at the same densities as in a).

4. Densities of states $N(E)$ obtained from the gaussian model of the nanotube distribution in ${\bf R}$ space, with the 
parameters  $\sigma=0.1$ K (full curve) and $\sigma=0.3$ K (dashed curve).

5. Low $T$ heat capacity obtained with the gaussian model, with  $\sigma = 0.1$ K and number densities $N=10^{-5}\AA^{-3}$, $N=10^{-6}\AA^{-3}$ and  $N=10^{-7}\AA^{-3}$ (from top to bottom).

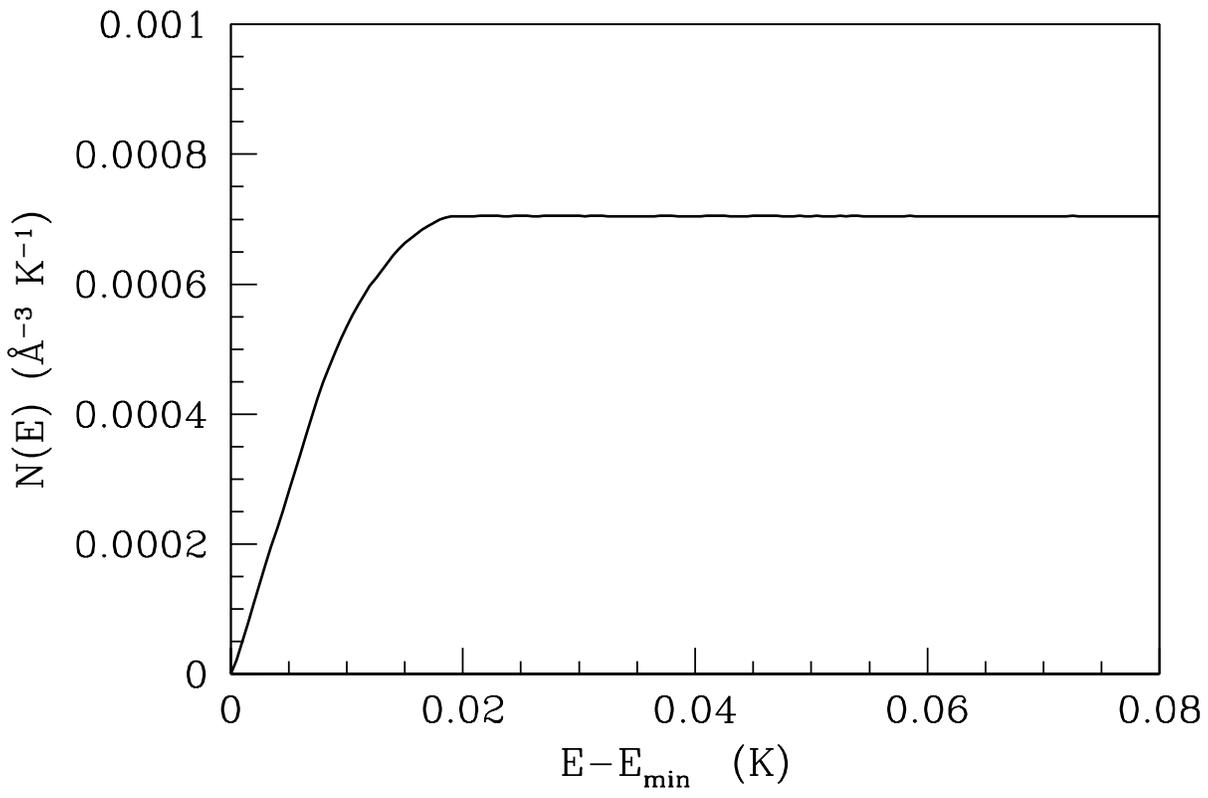
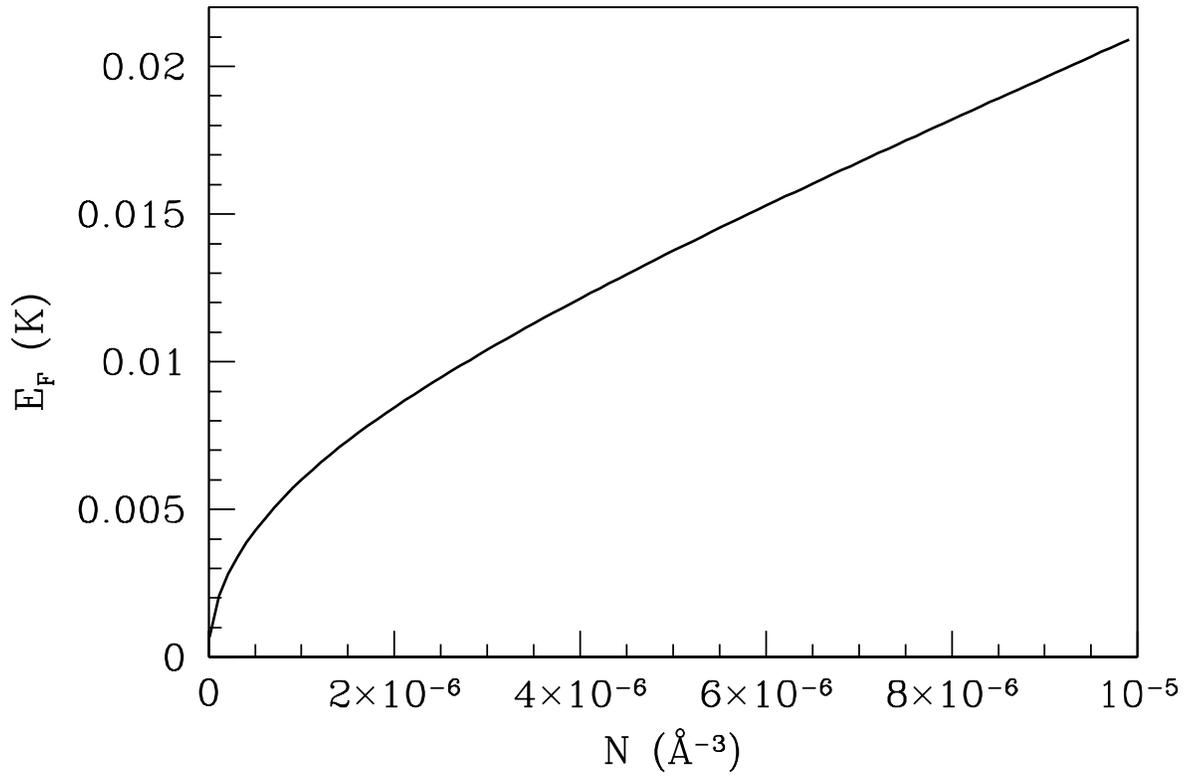
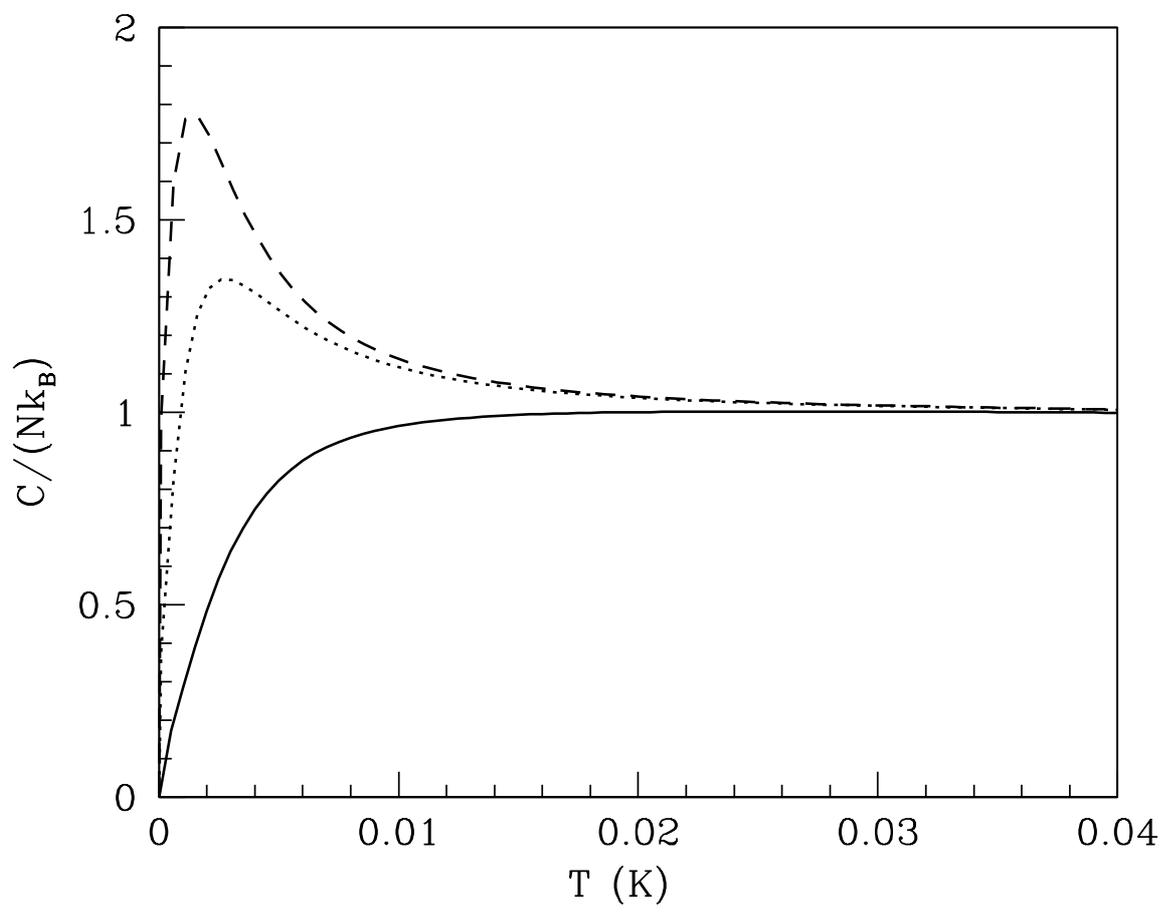
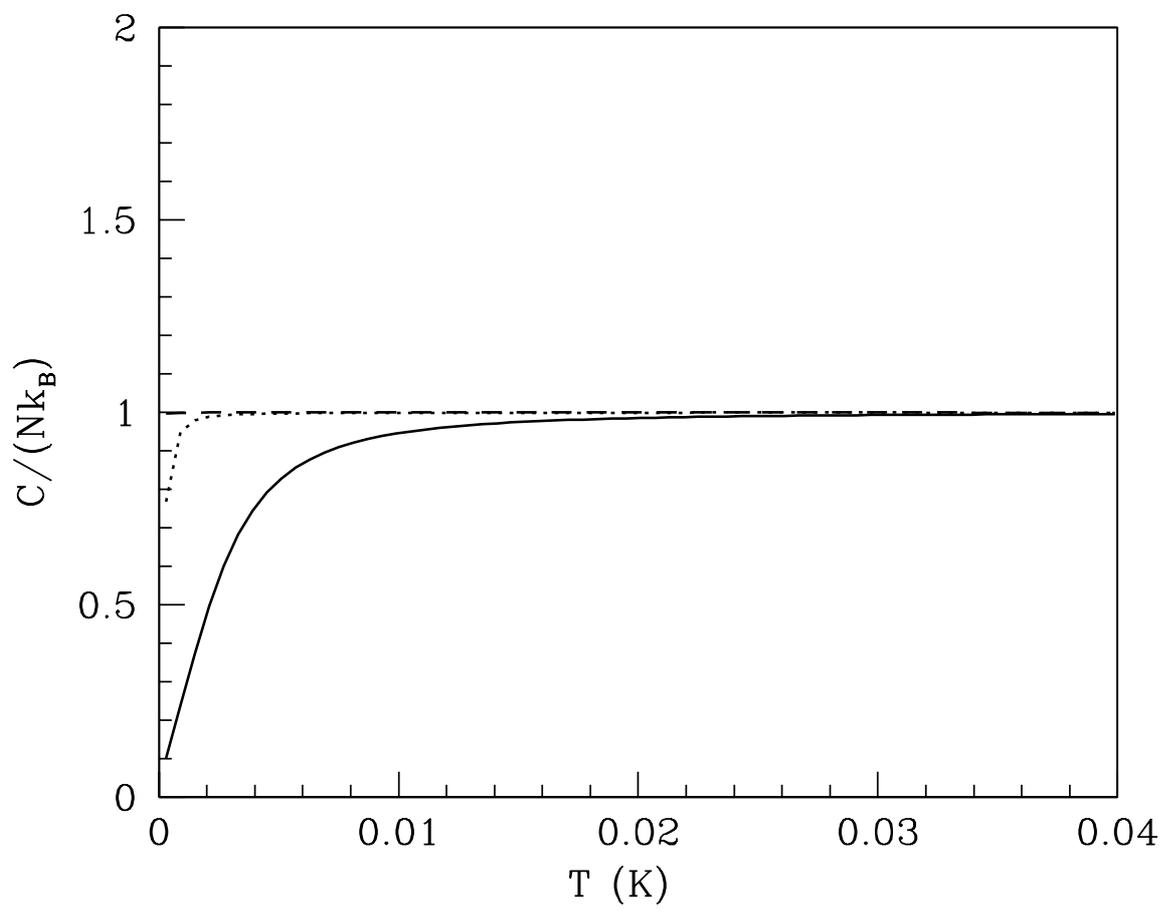
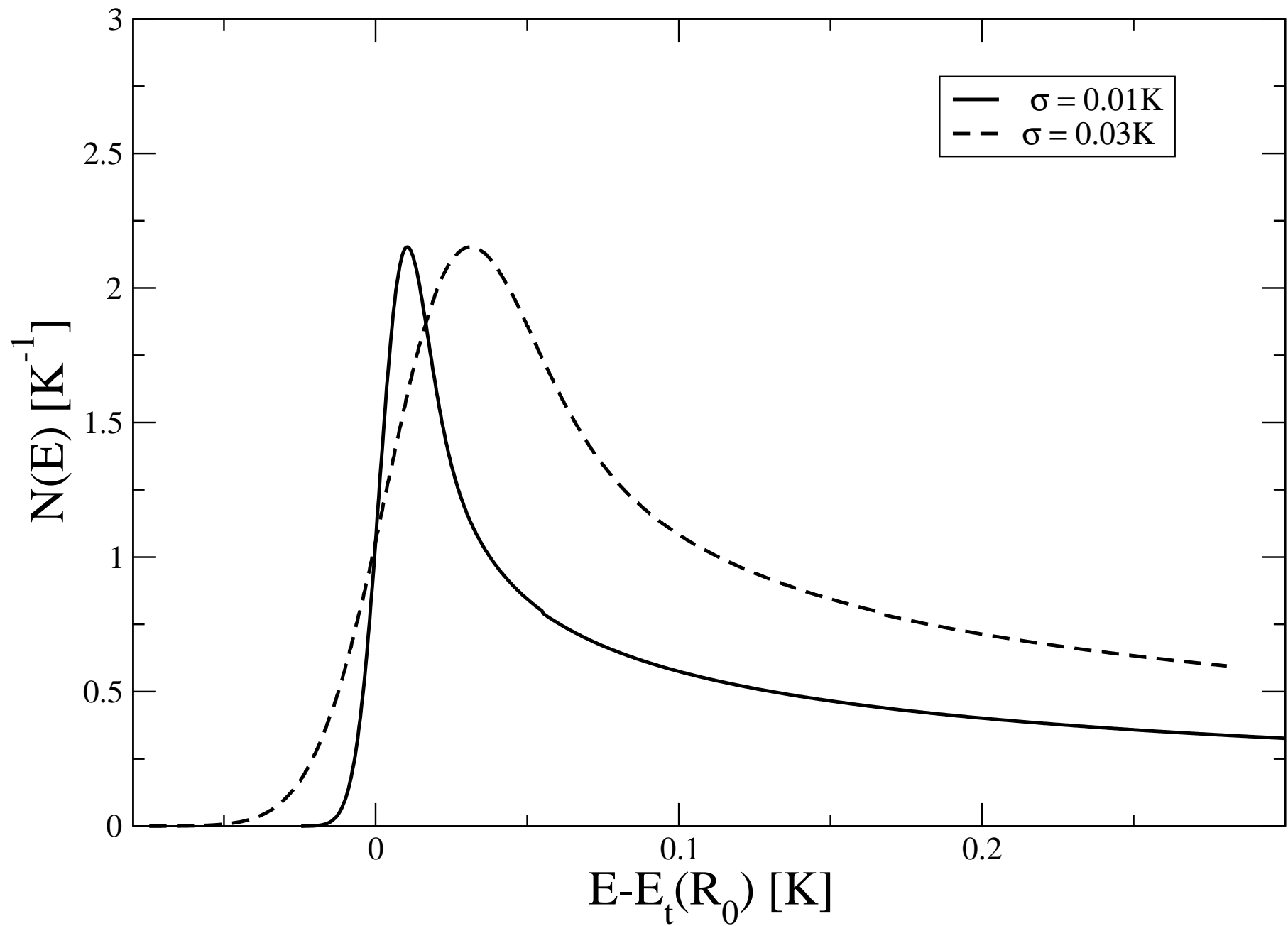
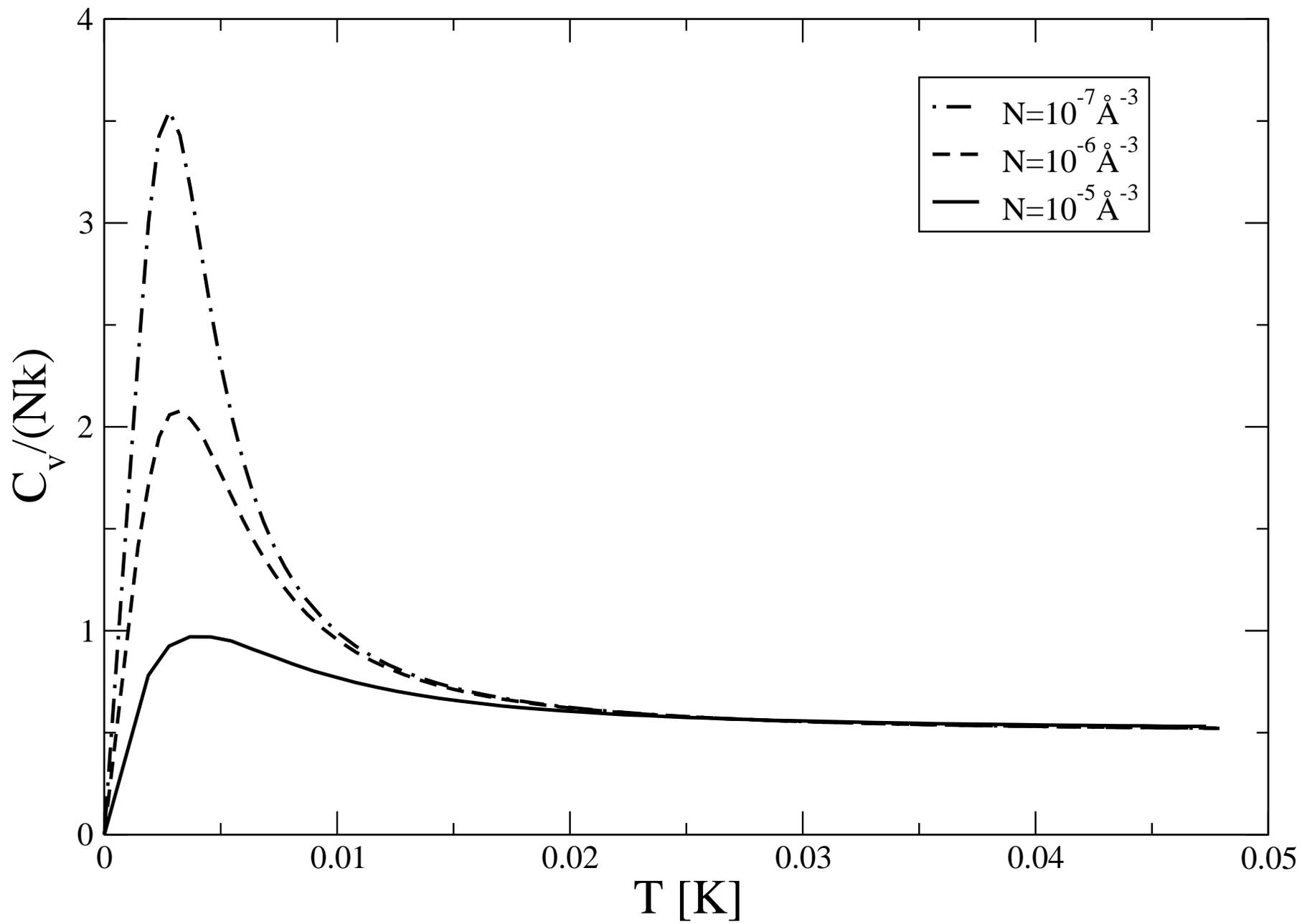
\end{document}